\documentclass[12pt]{article}
\usepackage[dvips]{epsfig}

\begin{document}

\author{A. de Souza Dutra$^{a,b}$\thanks{%
E-mail: dutra@feg.unesp.br}  and A. C. Amaro de Faria
Jr.$^{b}$ \\
$^{a}$Abdus Salam ICTP, Strada Costiera 11, Trieste, I-34100 Italy.\\
$^{b}$UNESCO-Campus de Guaratinguet\'{a}-DFQ\thanks{ Permanent
Institution}\\
Departmento de F\'{\i}sica e Qu\'{\i}mica\\
12516-410 Guaratinguet\'{a} SP Brasil}
\title{{\LARGE Expanding the class of general exact solutions for interacting two field kinks}}
\maketitle

\begin{abstract}
In this work we extend the range of applicability of a method
recently introduced where coupled first-order nonlinear equations
can be put into a linear form, and consequently be solved
completely. Some general consequences of the present extension are
then commented.

\end{abstract}

\newpage

Despite of being less common, the systems governed by nonlinear
equations are of great importance and, nowadays, there is a
growing interest in dealing with such systems \cite{witten} -
\cite{fosco}. Unfortunately, as a consequence of the nonlinearity,
in general we lose the capability of getting the complete
solutions. Here we extend a demonstration \cite{plb05} that is
profitable for those systems in 1+1 dimensions, whose the
second-order differential equations can be reduced to the solution
of corresponding first-order equations, the so called Bolgomol
'nyi-Prasad-Sommerfield (BPS) topological solitons \cite{BPS}. In
this approach one can obtain a differential equation relating the
two coupled fields which, once solved, leads to the general orbit
connecting the vacua of the model. In fact, the ``trial and error
'' methods historically arose as a consequence of the intrinsic
difficulty of getting general methods of solution for nonlinear
differential equations. About two decades ago, Rajaraman
\cite{rajaraman} introduced an approach of this nature for the
treatment of coupled relativistic scalar field theories in 1+1
dimensions. His procedure was model independent and could be used
for the search of solutions in arbitrary coupled scalar models in
1+1 dimensions. However, the method is limited in terms of the
generality of the solutions obtained and is convenient and
profitable only for some particular, but important, cases
\cite{boya}. Some years later, Bazeia and collaborators
\cite{bazeia0} applied the approach developed by Rajaraman to
special cases where the solution of the nonlinear second-order
differential equations are equivalent to the solution of
corresponding first-order nonlinear coupled differential
equations. By the way, Bazeia and collaborators wisely applied
their solution to a variety of natural systems, since polymers up
to domain walls. Last year \cite{plb05}, one of us showed that
many of the systems studied in \cite{bazeia0}-\cite{bazeia4} can
be mapped into a first-order linear differential equation and, as
a consequence, can be solved in order to get the general solution
of the system. In this work we are going to show that one can
extend the range of applicability of that procedure, by taking
into account situations where the potential of the system can not
be written in terms of a superpotential as it is usually supposed
to be necessary \cite{bazeia0}-\cite{bazeia4}. After that, we
trace some comments about the consequences, and possible
applications, coming from these general solutions.

In order to deal with the problem, following the usual procedure
to get BPS \cite{BPS} solutions for nonlinear systems, one usually
particularizes \cite{bazeia0}-\cite{bazeia4} the form of the
Lagrangian density
\begin{equation}
L=\frac{1}{2}\left( \partial _{\mu }\phi \right) ^{2}+\frac{1}{2}\left(
\partial _{\mu }\chi \right) ^{2}-V\left( \phi ,\chi \right) ,
\end{equation}

\noindent by imposing that the potential must be written in terms
of a superpotential $W\left(\phi,\chi\right)$ such that
\begin{equation}
V\left( \phi ,\chi \right) =\frac{1}{2}\left( \frac{\partial W\left( \phi
,\chi \right) }{\partial \phi }\right) ^{2}+\frac{1}{2}\left( \frac{\partial
W\left( \phi ,\chi \right) }{\partial \chi }\right) ^{2}.
\end{equation}

Then, the energy functional of the static configurations can be
calculated straightforwardly, giving
\begin{equation}
E_{B}=\frac{1}{2}\int_{-\infty }^{\infty }dx\left[ \left( \frac{d\phi }{dx}%
\right) ^{2}+\left( \frac{d\chi }{dx}\right) ^{2}+\,W_{\phi }^{2}+\,W_{\chi
}^{2}\right] ,
\end{equation}

\noindent which leads to
\begin{equation}
E_{B}=\frac{1}{2}\int_{-\infty }^{\infty }dx\left[ \left( \frac{d\phi }{dx}%
-W_{\phi }\right) ^{2}+\left( \frac{d\chi }{dx}-W_{\chi }\right)
^{2}+2 \,\left( W_{\chi }\frac{d\chi }{dx}+W_{\phi }\frac{d\phi
}{dx} \right) \right] ,
\end{equation}

\noindent and finally, considering the minimal energy field
configuration, one gets

\begin{equation}
E_{B}=|W\left( \phi _{j},\chi _{j}\right) -W\left( \phi _{i},\chi
_{i}\right) |,  \label{eq3}
\end{equation}

\noindent where $\phi _{i}$ and $\chi _{i}$ are the are the $i-th$ vacuum
state of the model \cite{bazeia1.5}.

In this case, one can easily see that solutions with minimal energy of the
second-order differential equation for the static solutions in 1+1
dimensions, can be solved through the corresponding first-order coupled
nonlinear equations
\begin{equation}
\frac{d\phi }{dx}=W_{\phi }\left( \phi ,\chi \right) \,,\,\frac{d\chi }{dx}%
=W_{\chi }\left( \phi ,\chi \right) ,  \label{eq1}
\end{equation}

\noindent where $W_{\phi }\equiv \frac{\partial W}{\partial \phi }$ and $%
W_{\chi }\equiv \frac{\partial W}{\partial \chi }$. It is
interesting to remark that the BPS solutions settle into vacuum
states asymptotically. In other words, the vaccum states act as
implicit boundary conditions of the BPS equations.

In this work we are going to expand the class of nonlinear models
with exact solutions by relaxing the condition over the potential
to a less restrictive one. Namely we uses that

\begin{equation}
V\left( \phi ,\chi \right) =\frac{1}{2}\left( A\left( \phi ,\chi
\right) \right) ^{2}+\frac{1}{2}\left( B\left( \phi ,\chi \right)
\right) ^{2}.
\end{equation}

\noindent As a consequence the energy functional can then be
written as

\begin{equation}
E_{B}=\frac{1}{2}\int_{-\infty }^{\infty }dx\left[ \left( \frac{d\phi }{dx}%
-A\right)^2+\left( \frac{d\chi }{dx}-B^{2}\right)^2 +2 \,\left( B
\frac{d\chi }{dx}+A \frac{d\phi }{dx} \right) \right] .
\end{equation}

Once again, it can be shown that the configuration which makes the
energy minimal is that where the fields obey the first-order
differential equations
\begin{equation}
\frac{d\phi }{dx}=A\left( \phi ,\chi \right) \,,\,\frac{d\chi }{dx}%
=B\left( \phi ,\chi \right) .  \label{eq2}
\end{equation}
\noindent This happens because, using the above equations we see
that the energy functional becomes simply
\begin{equation}
E_{B}=\int_{-\infty }^{\infty }dx\left[ \left( \frac{d\chi
}{dx}\right)^2+ \left( \frac{d\phi }{dx} \right)^2 \right] ,
\end{equation}
\noindent which is positive definite, so rendering the minimal
energy for the fields configuration obeying (\ref{eq2}).

Furthermore, in order to guarantee that the above equations obey
the correct second-order differential equations
\begin{equation}
\frac{d^2\phi}{dx^2}=V_{\phi }\left( \phi , \chi\right) ,\,\,\,
\frac{d^2\chi}{dx^2}=V_{\chi }\left( \phi , \chi\right) ,
\end{equation}
\noindent one must impose the following restriction over the, otherwise
arbitrary functions,
\begin{equation}
A_{\chi}\left( \phi , \chi\right)=B_{\phi}\left( \phi , \chi\right). \label{AB}
\end{equation}

Apart from this constraint and presenting a computationally more
involved calculation of the topological energy, this more general
situation can now be studied by using, for instance the usual
trial-orbit approach \cite{bazeia0}-\cite {bazeia4} or, by a
direct solution of the orbit equation as done in \cite{plb05}. In
this work we choose the second line of thinking. The idea is to
write the following equation
\begin{equation}
\frac{d\phi }{A}=dx=\frac{d\chi }{B},
\end{equation}

\noindent where the spatial differential element is a kind of invariant. So,
one obtains that
\begin{equation}
\frac{d\phi }{d\chi }=\frac{A\left( \phi ,\chi \right)}{B\left( \phi ,\chi \right)}.  \label{eqm}
\end{equation}

This last equation is, in general, a nonlinear differential equation
relating the scalar fields of the model. Now, if one is able to solve it
completely, the function $\phi \left( \chi \right) $ can be used to
eliminate one of the fields, so rendering the equations (\ref{eq1})
uncoupled. Finally, this uncoupled first-order nonlinear equation can be
solved in general, even if numerically.

At this point it is interesting to deal with a concrete example, in order to
better expose the advantage of the present approach. Let us use, for instance
the following functions $A$ and $B$, as the potential defining ones,
\begin{eqnarray}
A\left( \phi ,\chi \right)&\equiv & \frac{\alpha_1}{2} \phi^2+\frac{\beta_1}{2}
\chi^2+\gamma_1\,\phi\,\chi+\delta_1\,\phi+\epsilon_1\,\chi+\zeta_1 \nonumber \\
 \\
B\left( \phi ,\chi \right)&\equiv & \frac{\alpha_2}{2} \phi^2+\frac{\beta_2}{2}
\chi^2+\gamma_2\,\phi\,\chi+\delta_2\,\phi+\epsilon_2\,\chi+\zeta_2 . \nonumber
\end{eqnarray}

By requiring that the condition (\ref{AB}) over the $A$ and $B$
functions be kept, we get the constraints:
\begin{equation}
\beta_1 = \gamma_2 ;\, \gamma_1 = \alpha_2 ; \,\epsilon_1 =
\delta_2 .
\end{equation}

If one try to construct a superpotential from the above functions
$A$ and $B$, in the spirit which it is usually done, one concludes
that it is not possible to do so unambiguously, predominantly due
to the difference between the one-field dependent terms. For
instance in the above illustrative example one would obtain the
following two alternative superpotentials

\begin{equation}
W_1\equiv \int{A\,d\phi}; \,\,\,\,\, W_2\equiv \int{A\, d\chi},
\end{equation}

\noindent leading respectively to the following first-order BPS equations
\begin{equation}
\frac{d\phi }{d\chi }=\frac{A\left( \phi ,\chi \right)}{\left[ B\left(
\phi , \chi \right)-\left(\frac{\beta_2}{2}
\chi^2+\epsilon_2\,\chi+\zeta_2 \right)\right] },
\end{equation}
\noindent and

\begin{equation}
\frac{d\phi }{d\chi }=\frac{\left[A\left( \phi ,\chi \right)-\left(
\frac{\alpha_1}{2} \phi^2+\delta_1\,\phi+\zeta_1   \right)\right] }{ B\left(
\phi , \chi \right)}.
\end{equation}

So, as we can see from the above, one can at least extend the
usual BPS procedure through introduction of arbitrary functions
depending on only one of the interacting fields. In order to have
an exact explicit example, we further restrict our parameters
choice by putting $\gamma_1=\beta_2=\delta_1=
\epsilon_1=\epsilon_2=\zeta_2=0$, which leaves us with

\begin{equation}
\frac{d\phi }{d\chi }=\frac{A\left( \phi ,\chi \right)}{B\left( \phi ,\chi \right)}=
\frac{\frac{\alpha_1}{2} \phi^2+\frac{\beta_1}{2}\,\chi^2+\zeta_1}{\beta_1\,\phi\,\chi} ,
\end{equation}

\noindent which, after defining the field $\sigma\equiv
\phi^2+\frac{2}{\alpha_1} \zeta_1 $, leads us to the following
linear differential equation \cite{plb05},

\begin{equation}
\frac{d\sigma}{d\chi}=\frac{\alpha_1}{\beta_1\,\chi}\sigma+\beta_1\,\chi,
\end{equation}

\noindent and finally to the following exact orbit equation

\begin{equation}
\phi^2 = c_0
\chi^{\frac{\alpha_1}{\beta_1}}+\frac{\beta_{1}^{2}}{2\,\beta_1-\alpha_1
}\chi^2 -\frac{2\,\zeta_1}{\alpha_1},
\end{equation}

\noindent with $c_0$ being the arbitrary integration constant. In
this case the system potential is given by
\begin{equation}
V\left(\phi,\chi\right)=\frac{\alpha_1^2}{8}\phi^4+\frac{\alpha_1\,\zeta_1}{2}\phi^2
+\frac{\beta_1^2}{8}\chi^4+\frac{\beta_1\,\zeta_1}{2}\chi^2+\frac{\beta_1\left(
2\,\beta_1+\alpha_1 \right)}{4}\phi^2\chi^2+\frac{\zeta_i^2}{2},
\end{equation}
\noindent which is to be compared here with the one discussed by
Boya and Casahorran some years ago \cite{boya}, and can be
written, identifying their fields and ours through
$\sigma\equiv\phi$ as $\rho\equiv\chi$, as
\begin{equation}
V_{BC}\left(\phi,\chi\right)=\frac{1}{4}\phi^4-\frac{1}{2}\phi^2
+\frac{\lambda}{4}\chi^4+\frac{\left(
f-d\right)}{2}\chi^2+\frac{d}{2}\phi^2\chi^2+\frac{1}{4},
\end{equation}
\noindent from which we can map the constants of the two models in
order to see, if possible, in what situation they can me mapped.
From an inspection of the above potentials we conclude that
$\alpha_1=\sqrt{2}$, $\zeta_1=-\frac{1}{\sqrt{2}}$,
$\beta_1=\pm\,\sqrt{2\,\lambda}$, and also the following
restriction over the original parameters in $V_{BC}$ should be
taken into account,
\begin{equation}
d=\sqrt{\lambda}\left( 2\,\sqrt{\lambda}\pm
1\right),\,\,f=2\,\lambda ,
\end{equation}

\noindent which implies into a constraint over the these two last
parameters such that $d= f \pm \frac{\sqrt{f}}{2}$.

At this point it is important to stress that a system like that
studied in \cite{boya} can not have their solutions reached by the
BPS usual procedure \cite{bazeia0}-\cite{bazeia4}. On the other
hand, in their work \cite{boya} Boya and Casahorran have made a
careful analysis of the soliton stability, showing under what
circumnstances they are stable under linear perturbations, which
now can be extended to our case, at least when the above
appropriate constraint holds. Furthermore, now in a similar
fashion to what has been done in the case of superpotential
procedure to get topological configurations \cite{plb05}, we can
show that there are further solutions to those obtained in
\cite{boya}, at least when the above mentioned restrictions are
kept. Those mentioned solutions have the special property of being
kink-like for both fields, in contrast with happens with the
already known ones \cite{boya}. An important remark at this point,
is that here the solutions are by no means obtained through a
trial and error approach, instead, they are exactly solved.

Let us consider now a bit more general model, where
$\gamma_1=\beta_2=\delta_1= \epsilon_2=\zeta_2=0$. By using the
same field transformation than in the previous case we get the
following linear differential equation
\begin{equation}
\frac{d\sigma}{d\chi}=\left[\frac{\left( \alpha_1 \sigma+
\beta_1\,\chi^2+2\,\epsilon_1\chi\right)}{\left(\beta_1\chi+\epsilon_1\right)}\right],
\end{equation}
which, after solved, leaves us with the exact orbit equation,
connecting the two fields,
\begin{equation}
\sigma\left(
\chi\right)=\left[\frac{2\,\epsilon_1^2+2\,\alpha_1\,\epsilon_1\,\chi+\alpha_1\,\beta_1\,\chi^2}{\alpha_1\left(
2\,\beta_1-\alpha_1\right)}\right]
+c_0\left(\beta_1\,\chi+\epsilon_1\right)^{\frac{\alpha_1}{\beta_1}},
\end{equation}

\noindent for $\alpha_1\neq 2\,\beta_1$, and
\begin{equation}
\sigma\left(
\chi\right)=\frac{\epsilon_1^2}{2\,\beta_1^2}+c_1\,\left(\epsilon_1+\beta_1\,\chi\right)^2+
\frac{\left(\epsilon_1+\beta_1\,\chi\right)^2}{\beta_1^2}ln\left(\epsilon_1+\beta_1\,\chi\right),
\end{equation}

\noindent for $\alpha_1 = 2\,\beta_1$; $c_0$ and $c_1$ are
arbitrary integration constants. From now on, we substitute these
solutions in one of the equations (\ref{eq2}), and solve it, so
obtaining a general solution for the system. In general it is not
possible to solve $\chi $ in terms of $\phi $ from the above
solutions, but the contrary is always granted. Here we will
substitute $\phi \left( \chi \right) $ in the equation for the
field $\chi $. However, for the case $\alpha_1 = 2\,\beta_1$ it is
very hard to get an exact solution and we will, as usually done in
this type of situation \cite{plb05} to the deal with the case
where $\alpha_1\neq 2\,\beta_1$. By doing the described
substitution, one gets
\begin{equation}
\frac{d\chi }{dx}=\pm \,\left(
\beta_1\,\chi+\epsilon_1\right)\,\sqrt{\frac{2\,\epsilon_1^2+2\,\alpha_1\,\epsilon_1\,\chi+\alpha_1\,\beta_1\,\chi^2}{\alpha_1\left(
2\,\beta_1-\alpha_1\right)}
+c_0\left(\beta_1\,\chi+\epsilon_1\right)^{\frac{\alpha_1}{\beta_1}}-\frac{2\,\zeta_1}{\alpha_1}}
. \label{tf1}
\end{equation}

In general we can not have an explicit solution for the above
equation, but one can verify numerically that the solutions are
always of the same classes. Notwithstanding, some classes of
solutions can be written in closed explicit forms. First of all,
we should treat the system when $c_{0}=0$, because in this
situation we can solve analytically the system for any value of
the parameters, apart from the case where $\alpha_1 = 2\,\beta_1$
as stated in above.

In order to  get an explicit exact solution in this situation, we
make a re-scaling in the spatial variable, such that $x=\beta_1\,
y$, and note that starting from the Ricatti equation
\begin{equation}
\frac{d\,\varphi}{dy}=a\,\left( \varphi^2+\frac{B^2}{4\,
A}-C\right) , \label{ricatti}
\end{equation}
\noindent and making the field transformation $\varphi\equiv
\sqrt{A\,\chi^2+B\,\chi+C}$, we finish with
\begin{equation}
\frac{d\,\chi}{dy}=a\,\left( \chi+\frac{B}{2\,
A}\right)\sqrt{A\,\chi^2+B\,\chi+C}. \label{oursystem}
\end{equation}

Now, performing the following identifications,
\begin{equation}
A\equiv\frac{\beta_1}{\left(
2\,\beta_1-\alpha_1\right)};\,\,B\equiv\frac{2\,\epsilon_1}{\left(
2\,\beta_1-\alpha_1\right)};\,\,C\equiv\frac{2}{\alpha_1}\left[
\frac{\epsilon_1^2}{\left(
2\,\beta_1-\alpha_1\right)}-\zeta_1\right],
\end{equation}

\noindent we are able to connect the equations (\ref{tf1}) and
(\ref{oursystem}), which allows us to obtain the solution we are
seeking for, just by using the solution of the Ricatti equation
(\ref{ricatti}), and substituting it in
\begin{equation}
\chi_{\pm}\equiv\frac{1}{2\, A}\left[ -\, B\pm\sqrt{B^2-4\,
A\left(C-\varphi^2\right)}\right],\,\, A\neq 0, \label{t1}
\end{equation}
\noindent or
\begin{equation}
\chi\equiv\frac{\left(\varphi^2-C\right)}{B},\,\, A=0,\,B\neq 0 .
\label{t2}
\end{equation}
\noindent In this second hypothesis, which is not allowed for the
set of parameters of the model we are considering, but which could
appear in another system, we must take into account that the
starting Ricatti equation should be written as
\begin{equation}
\frac{d\,\varphi}{dy}=\pm 2\,\left(
\varphi^2-C-\frac{\epsilon_1\,B}{\beta_1}\right) ,
\label{ricattib}
\end{equation}
\noindent and performing the transformation $\varphi\equiv
\sqrt{B\,\chi+C}$, we finish with
\begin{equation}
\frac{d\,\chi}{dy}=\pm\,\left(
\chi+\frac{\epsilon_1}{\beta_1}\right)\sqrt{B\,\chi+C}.
\label{oursystemb}
\end{equation}

Our final task is to obtain the solution of (\ref{ricatti}). In
fact this a very well known equation but, for the sake of
completeness, we map it too into a linear equation from which the
solution of our original coupled nonlinear first-order
differential equations will come after a sequence of mathematical
manipulations. For this we perform a final transformation
$\varphi\equiv\pm\sqrt{b}+\frac{1}{\Omega}$, where
$b\equiv\frac{B^2}{4\,A}-C$ ($A\neq0$), leaving us with
\begin{equation}
\frac{d\Omega}{dy}=a\left( 1\pm 2\,\sqrt{b}\,\Omega\right) ,
\end{equation}
\noindent from whose solution we obtain
\begin{equation}
\varphi\left( y\right)=\pm\sqrt{b}-\frac{2\,\sqrt{b}}{\pm
1-e^{-2\,a\,\sqrt{b}\,\left( y-y_0\right)}},
\end{equation}
\noindent with $y_0$ being the arbitrary integration constant. In
order to avoid a singularity in the solution, we must choose the
negative sign in the above solution. After that, it is not hard to
verify that the final function can be expressed as the one having
the usual kink profile,
\begin{equation}
\varphi\left( y\right) =\sqrt{b}\, tanh\left[
a\,\sqrt{b}\,\left(y-y_0\right)\right].
\end{equation}

However, once the field $\varphi$ appears quadratically in the
equations (\ref{t1}) and (\ref{t2}), the solution for this field
will never have a kink profile. It will always have a lump
behavior. This feature of the $c_0=0$ case was already observed in
\cite{plb05}, and here we confirm that this behavior is still
present in this more general case. Notwithstanding, we remark that
if $\epsilon_1\neq 0$, there is a remaining possibility in this
more general system, which were not present in the simpler case
studied in \cite{plb05}. For this case, we can look for solutions
in a special situation where the $\chi$ field equation can be
written in a simpler polynomial form. This happens provided that
$B\equiv\sqrt{A\,C}$, and in this case one gets
\begin{equation}
\frac{d\,\chi}{dy}=a\,\left(
\chi+\frac{\epsilon_1}{\beta_1}\right)\sqrt{A\,\chi^2+B\,\chi+C}=a\,\left(
\chi+\frac{B}{2\, A}\right)\left(\sqrt{A}\,\chi+\sqrt{C}\right),
\label{oursystem2}
\end{equation}
\noindent and finally
\begin{equation}
\frac{d\,\chi}{dy}=a\,\sqrt{A}\left(\chi'^2-b'\right),
\label{oursystem3}
\end{equation}
\noindent where
$\chi'\equiv\chi+\frac{1}{2}\left(\sqrt{\frac{C}{A}}+\frac{\epsilon_1}{\beta_1}\right)$
and
$b'\equiv\frac{1}{4}\left(\sqrt{\frac{C}{A}}-\frac{\epsilon_1}{\beta_1}\right)^2$,
\noindent with the kink-like solution
\begin{equation}
\chi=\sqrt{b'}\, tanh\left[ a\,\sqrt{A\,
b'}\,\left(y-y_0\right)\right]-\frac{1}{2}\left(\sqrt{\frac{C}{A}}+\frac{\epsilon_1}{\beta_1}\right).
\end{equation}

 It is not difficult to conclude that this last solution will lead
 to what we call type B kink \cite{plb05}, where both interacting fields $\chi$
 and $\phi$ present a kink profile, and the previous solution
 corresponds to that usually obtained in the literature
 \cite{bazeia0},\cite{bazeia1}-\cite{bazeia4}, what we call type A
 kink and is characterized by a lump form for the $\chi$ field and
 a kink form for the $\phi$ field.

 As observed in \cite{plb05}, for the case with $c_0\neq 0$ we
 can, at least for certain relations between the potential
 parameters, obtain analytical exact solutions for the type B kink solution. This can be
 achieved through a special fixing of the arbitrary integration
 parameter $c_0$. For instance, in the case of the model on the
 screen in this work, we could choose $\alpha_1=\beta_1$ and,
 after rearranging the terms in the equation (\ref{oursystem2}), conclude
 that the condition required in order to get a polynomial at the
 right-hand side of the equation for the field $\chi$, could be
 granted by imposing the following condition over $c_0$,
 \begin{equation}
 B+c_0\, \beta_1=\sqrt{A\left(C+c_0\,\epsilon_1\right)}.
 \end{equation}
 A similar analysis could be done for the case where
 $\alpha_1=4\,\beta_1$ which, despite of being more involved, lead
 us to the same conclusion about the possibility of generating
 type B kink through a convenient choice of $c_0$.

I this work we showed how to extend an approach recently
introduced \cite{plb05}, in order to enlarge the class of
nonlinear coupled field models which can have their solitonic
configurations exactly solved in $1+1$ dimensions. This allowed us
to, at last under certain special conditions, include models like
those studied by Boya and Casahorran \cite{boya} in this class. As
a bonus, we got a kink configuration which was not reported in
that work. Furthermore, it is important to assert that the class
of models which we were able to deal here could not be considered
through the usual procedure, based on the idea of having a
superpotential \cite{bazeia0}-\cite{bazeia4}. Finally we should
comment that similar models were very recently considered in the
literature, concerning the study of Lorentz breaking models
\cite{barreto,fabricio}.

\bigskip\bigskip

\noindent \textbf{Acknowledgments:} The author is grateful to CNPq
for partial financial support, and to Professor D. Bazeia for
introducing him to this matter. This work has been finished during
a visit within the Associate Scheme of the Abdus Salam ICTP.

\end{document}